\documentclass[conference,a4paper]{IEEEtran}

\usepackage[utf8]{inputenc} 
\usepackage[T1]{fontenc}
\usepackage{url}
\usepackage{ifthen}
\usepackage{cite}
\usepackage[cmex10]{amsmath} 

\usepackage{amsfonts}
\usepackage{accents}

\ifCLASSINFOpdf
\usepackage[pdftex]{graphicx}
\else
\usepackage[dvips]{graphicx}
\fi
\ifCLASSOPTIONcompsoc
\usepackage[caption=false,font=normalsize,labelfont=sf,textfont=sf]{subfig}
\else
\usepackage[caption=false,font=footnotesize]{subfig}
\fi

\usepackage{mathtools}
\usepackage{amssymb}
\usepackage{algorithm, algorithmic, float}
\usepackage{tikz}
\usetikzlibrary{automata, positioning}
\usetikzlibrary{backgrounds}
\usepackage{rotating}

\DeclarePairedDelimiter{\floor}{\lfloor}{\rfloor}

\newcommand{\mtx}{\mathbf}
\newcommand{\vct}{\mathbf}

\DeclareMathSymbol{\widehatsym}{\mathord}{largesymbols}{"62}
\newcommand\lowerwidehatsym{%
	\text{\smash{\raisebox{-1.3ex}{%
				$\widehatsym$}}}}
\newcommand\fixwidehat[1]{%
	\mathchoice
	{\accentset{\displaystyle\lowerwidehatsym}{#1}}
	{\accentset{\textstyle\lowerwidehatsym}{#1}}
	{\accentset{\scriptstyle\lowerwidehatsym}{#1}}
	{\accentset{\scriptscriptstyle\lowerwidehatsym}{#1}}
}

\newtheorem{theorem}{Theorem}
\newtheorem{lemma}{Lemma}

\begin{document}
	\title{
		Hierarchical Coding for Distributed Computing} 
	
	\author{%
		\IEEEauthorblockN{Hyegyeong Park, Kangwook Lee, Jy-yong Sohn, Changho Suh and Jaekyun Moon}
		\IEEEauthorblockA{School of Electrical Engineering, Korea Advanced Institute of Science and Technology (KAIST)\\
			Email: \{parkh, kw1jjang, jysohn1108, chsuh\}@kaist.ac.kr, jmoon@kaist.edu}
	}

	\maketitle
	
	\begin{abstract}
		Coding for distributed computing supports low-latency computation by relieving the burden of straggling workers. While most existing works assume a simple master-worker model, we consider a hierarchical computational structure consisting of groups of workers, motivated by the need to reflect the architectures of real-world distributed computing systems. In this work, we propose a \emph{hierarchical coding scheme} for this model, as well as analyze its decoding cost and expected computation time. Specifically, we first provide upper and lower bounds on the expected computing time of the proposed scheme. We also show that our scheme enables efficient parallel decoding, 
		thus reducing decoding costs by orders of magnitude over non-hierarchical schemes. When considering both decoding cost and computing time, the proposed hierarchical coding is shown to outperform existing schemes in many practical scenarios. 
	\end{abstract}
	
	
	\section{Introduction}
Enabling large-scale computations for big data analytics, distributed computing systems have received significant attention in recent years \cite{Dean12}.
	The distributed computing system divides a computational task to a number of subtasks, each of which is allocated to a different worker. 
	This helps reduce computing time by exploiting parallel computing options and thus enables handling of large-scale computing tasks.
	
In a distributed computing system, the ``stragglers'', which refers to the computing nodes that slow down in some random fashion due to a variety of factors, may increase the total runtime of the computing system.
	To address this problem, 
	the notion of \textit{coded computation} is introduced in \cite{Lee17TIT} where an $(n, k)$ maximum distance separable (MDS) code is employed to speed up distributed matrix multiplications.  
	The authors show that for linear computing tasks, one can design $n$ distributed computing tasks such that \emph{any} $k$ out of $n$ tasks suffice to complete the assigned task. 
	Since then, coded computation has been applied to a wide variety of task scenarios such as matrix-matrix multiplication~\cite{Lee17ISIT, Yu17}, distributed gradient computation~\cite{Tandon17,Halbawi17,Raviv17,Charles17}, convolution~\cite{Dutta17}, Fourier transform~\cite{Yu17FFT} and matrix sparsification~\cite{Dutta16, Suh17}.

	
	While the idea of coded computation has been studied in various settings, existing works have not taken into account the underlying 
	\emph{hierarchical} nature of practical distributed systems~\cite{Dean08, Ahmad14,Vahdat10}. In modern distributed computing systems,
	each group of workers is collocated in the same rack, which contains a Top of Rack (ToR) switch, and cross-rack communication is available only via these ToR switches. Surveys on real cloud computing systems show that cross-rack communication through the ToR switches is highly unstable due to the limited bandwidth, whereas intra-rack communication is faster and more reliable \cite{Ahmad14,Vahdat10}.
	A natural question is whether one can devise a coded computation scheme that exploits such hierarchical structure.

	\begin{figure}[t]
		\centering
		\includegraphics[width=1\linewidth]{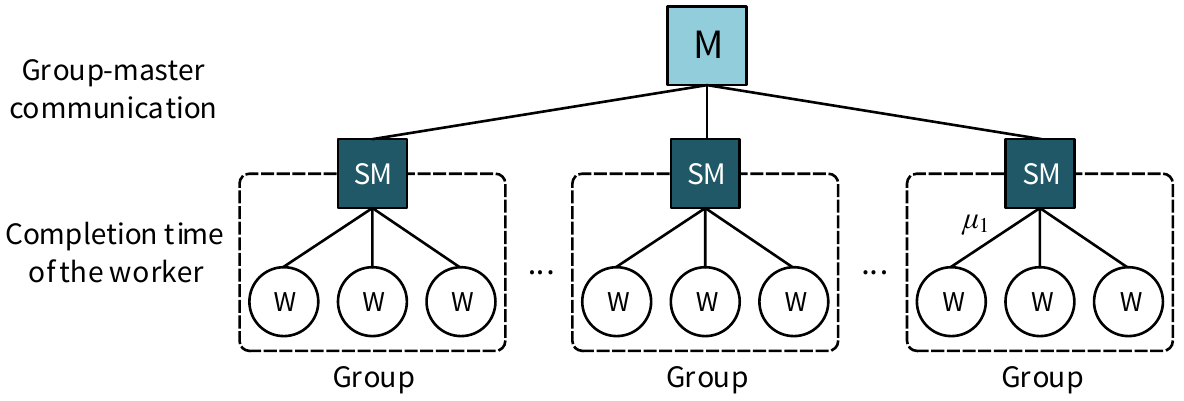}
		\caption{Illustration of the hierarchical computing system}
		\label{fig: hier}
		\vspace{-0.15in}
	\end{figure} 
	
	\subsection{Contribution}
	In this work, we first model a distributed computing system with a tree-like hierarchical structure illustrated in Fig. \ref{fig: hier}, which is inspired by the practical computing systems in ~\cite{Dean08, Ahmad14,Vahdat10}. 
	The workers (denoted by ``W'') are divided into groups, each of which has a submaster (denoted by ``SM''). Each submaster sends the computational result of its group to the master (denoted by ``M'').
	The suggested model can be viewed as a generalization of the existing non-hierarchical coded computation.
	
\begin{figure}[t]
		\centering
		\includegraphics[width=1\linewidth]{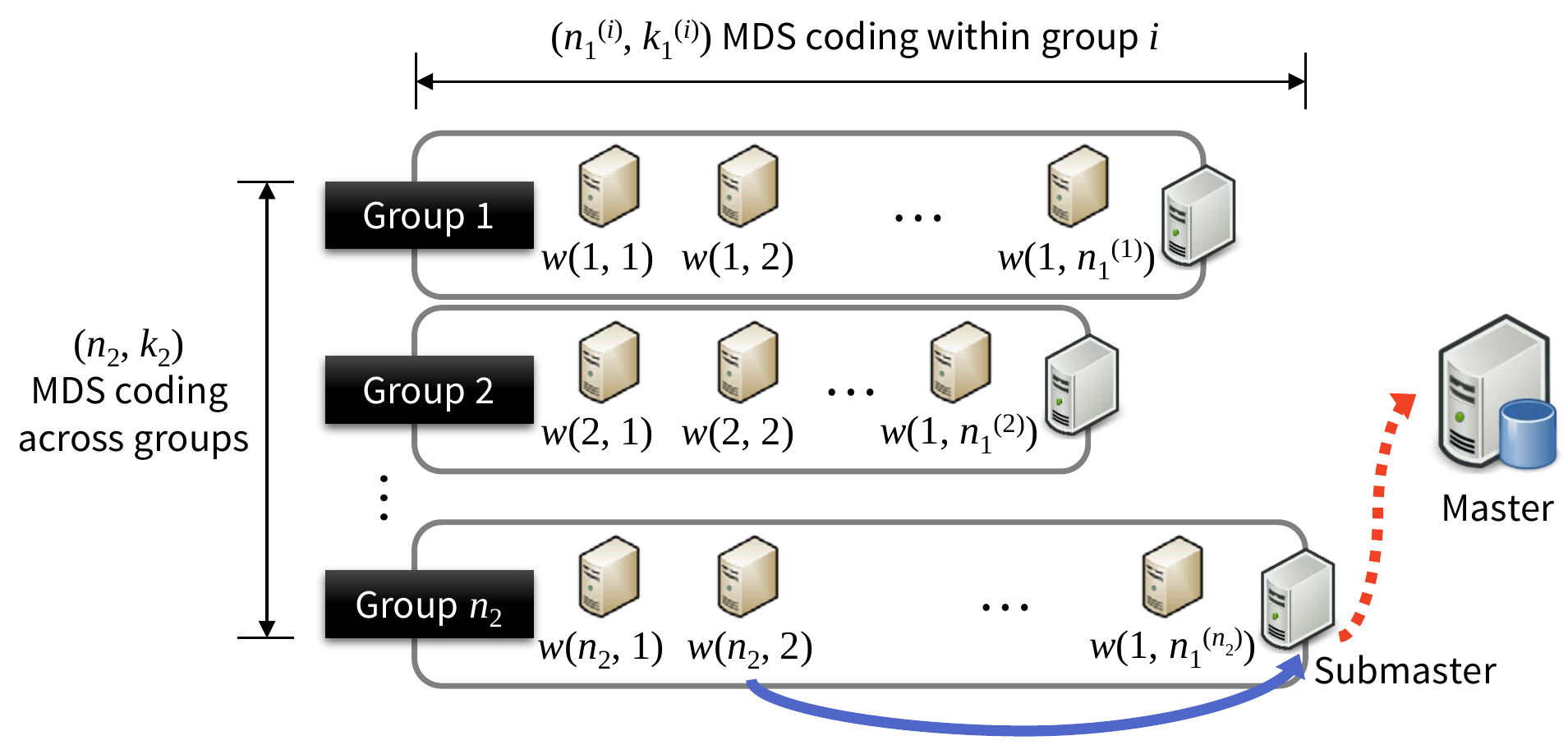}
		\caption{Illustration of the proposed coding scheme applied to the hierarchical computing system. An $(n_1^{(i)}, k_1^{(i)})$ MDS code is employed within group $i$, and an $(n_2, k_2)$ MDS code is applied across the groups. $w(i, j)$ denotes worker $j$ in group $i$.}
		\label{fig: cluster}
		\vspace{-0.1in}
	\end{figure}

	In this framework, we propose a hierarchical coding scheme which employs an $(n_1^{(i)}, k_1^{(i)})$ MDS code within group $i$ and another $(n_2, k_2)$ outer MDS code across the groups as depicted in Fig. \ref{fig: cluster}. 
We also develop a parallel decoding algorithm which exploits the concatenated code structure and allows low complexity.
Moreover, we analyze the latency performance of our proposed solution. 
It turns out that the latency performance of our scheme cannot be analyzed via simple order statistics as in other existing schemes.
Here we resort to find lower and upper bounds on the average latency performance:
Our upper bound relies on concentration inequalities, and our lower bound is obtained via constructing and analyzing an auxiliary Markov chain (to be detailed later).

	\subsection{Related Work}
	Previous works on coded computation have rarely considered the inherent hierarchical structure of most real-world systems. 
	Whereas a very recent work \cite{Gupta17} deals with the multi-rack computing system reflecting
	imbalance between intra- and cross-rack communications,
	it is based on the settings of the coded MapReduce architecture
	which do not include general linear computation tasks that we focus on in this work. Another distinction is that the analysis of \cite{Gupta17} includes only the cross-rack redundancy whereas our analysis considers both intra- and cross-group coding. 
	
	\subsection{Notations}
	We use boldface uppercase letters for matrices and boldface lowercase letters for vectors. The transpose of a matrix $\mtx{A}$ is denoted by $\mtx{A}^T$. 
	For a matrix $\mtx{A}$ satisfying $\mtx{A}^T = [\mtx{A}_{1}^T \,\, \mtx{A}_2^T]$, we write $\mtx{A} = [\mtx{A}_1 ; \mtx{A}_2]$.
	For a positive integer $n$, the set $\{1, 2, \dots, n\}$ is denoted by $[n]$. The $j\textsuperscript{th}$ worker in group $i$ is represented by $w(i,j)$ for $i \in [n_2]$ and $j \in [n_1^{(i)}]$.
	The symbol $\floor*{r}$ indicates the largest integer less than or equal to a real number $r$.
	
	\section{Hierarchical Coded Computation}\label{sec: model}

	\subsection{Proposed Coding Scheme}	
	
	Consider a matrix-vector multiplication task, i.e., computing $\mtx{A} \vct{x}$ for a matrix $\mtx{A} \in \mathbb{R}^{m \times d}$ and a vector $\vct{x} \in  \mathbb{R}^{d \times 1}$.
	The input matrix $\mtx{A}$ is split into $k_2$ submatrices as $\mtx{A} = [\mtx{A}_1;  \mtx{A}_2;  \dots ; \mtx{A}_{k_2}]$, where $\mtx{A}_i \in \mathbb{R}^{\frac{m}{k_2} \times d}$ for $i \in [k_2]$. Here we assume that $m$ is divisible by $k_2$ for simplicity. Then, we apply an $(n_2, k_2)$ MDS code to set $\{\mtx{A}_i\}_{i \in [k_2]}$ in obtaining $\{\widetilde{\mtx{A}}_i\}_{i\in [n_2]}$. Then, each coded matrix $\widetilde{\mtx{A}}_{i}$ is further divided into $k_1^{(i)}$ submatrices as $\widetilde{\mtx{A}}_{i} = [\widetilde{\mtx{A}}_{i,1};  \widetilde{\mtx{A}}_{i,2};  \dots;
	\widetilde{\mtx{A}}_{i,k_1^{(i)}}]$
	where $\widetilde{\mtx{A}}_{i,j} \in \mathbb{R}^{\frac{m}{k_1^{(i)}k_2} \times d}$ for $j \in [k_1^{(i)}]$ and $m$ divisible by $k_1^{(i)}k_2$. Afterwards, for each  $i\in[n_2]$, we apply an $(n_1^{(i)}, k_1^{(i)})$ MDS code to set $\{\widetilde{\mtx{A}}_{i,j}\}_{j \in [k_1^{(i)}]}$ to obtain $\{\fixwidehat{\mtx{A}}_{i,j}\}_{j\in [n_1^{(i)}]}$.
	Then, for each $i\in[n_2]$ and $j\in[n_1^{(i)}]$, worker $w(i,j)$ computes $\fixwidehat{\mtx{A}}_{i,j}\vct{x}$. 
Fig. \ref{fig: cluster} illustrates the proposed coding scheme for the hierarchical computing system with a different number of workers in each group.
In the case of $n_1^{(i)} = n_1$ and $k_1^{(i)} = k_1$ for all $i \in [n_2]$, we will refer this coding scheme as $(n_1,k_1) \times (n_2, k_2)$ coded computation.

	\begin{figure}[t]
		\centering\includegraphics[width=1\linewidth]{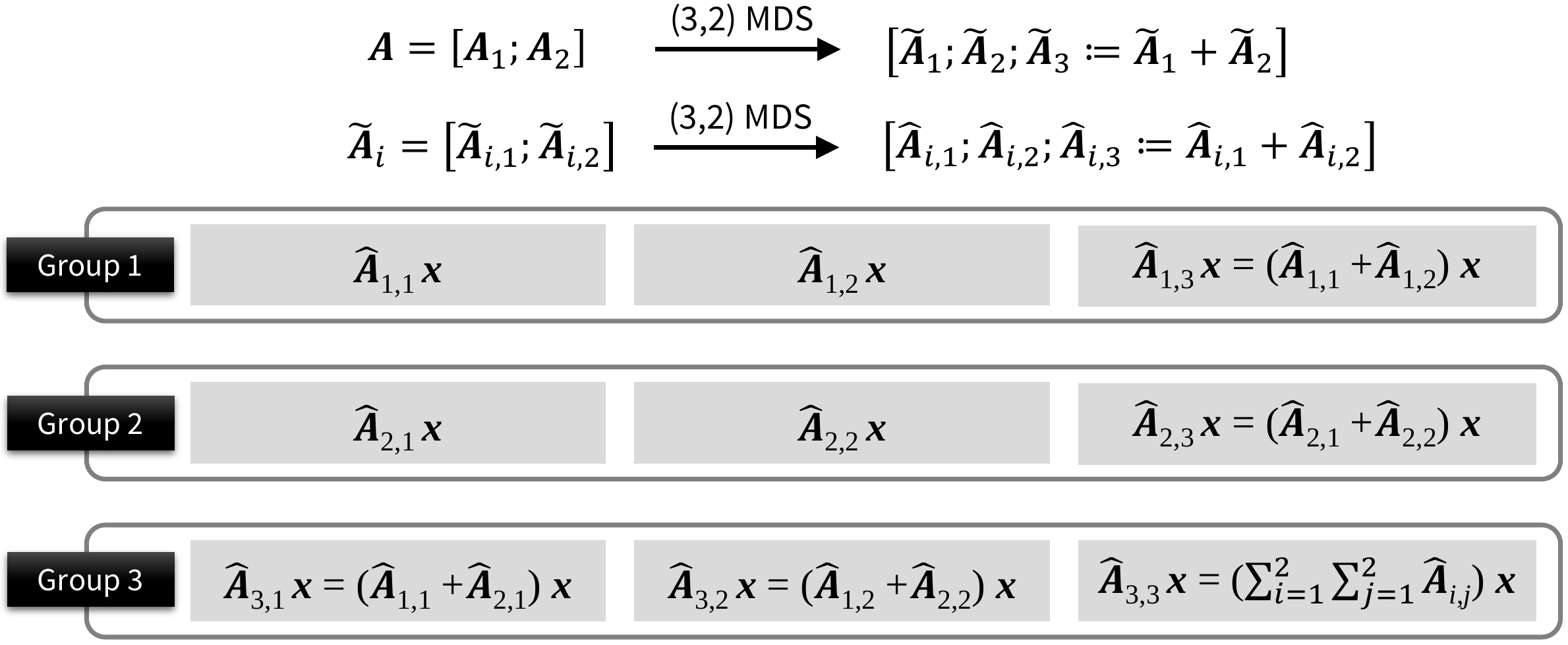}
		\caption{Allocation of the computational task to workers in a $(3, 2)\times(3, 2)$ coded computation}
		\label{fig: alloc}
		\vspace{-0.1in}
	\end{figure}
	We present our code in Fig.~\ref{fig: alloc} via a toy example.
	In this example, $(n_1,k_1)=(n_2,k_2)=(3,2)$.
	That is, the input matrix $\mtx{A}  =[\mtx{A}_1; \mtx{A}_2]$ is encoded via $(n_2,k_2)=(3,2)$ MDS code, yielding $[\widetilde{\mtx{A}}_1;
	\widetilde{\mtx{A}}_2; \widetilde{\mtx{A}}_1+\widetilde{\mtx{A}}_2]$. 
	Afterwards, the matrix $\widetilde{\mtx{A}}_i = [\widetilde{\mtx{A}}_{i,1}; \widetilde{\mtx{A}}_{i,2}]$ is encoded via an $(n_1,k_1)=(3,2)$ MDS code, producing $[\fixwidehat{\mtx{A}}_{i,1}; \fixwidehat{\mtx{A}}_{i,2}; \fixwidehat{\mtx{A}}_{i,1} +\fixwidehat{\mtx{A}}_{i,2}] $.
	For notational simplicity, we define $\widetilde{\mtx{A}}_{3} = \widetilde{\mtx{A}}_{1} + \widetilde{\mtx{A}}_{2}$ and
	$\fixwidehat{\mtx{A}}_{i,3} = \fixwidehat{\mtx{A}}_{i,1} + \fixwidehat{\mtx{A}}_{i,2}$.
	For $i, j\in \{1, 2, 3\}$, worker $w(i, j)$ computes $\fixwidehat{\mtx{A}}_{i,j}\vct{x}$.
	Note that group $i$ is assigned a subtask with respect to $\widetilde{\mtx{A}}_{i}$.

	We now describe the decoding algorithm for our proposed coding scheme. 
	When a worker completes its task, it sends the result to its submaster. 
	With the aid of the $(n_1,k_1)$ MDS code, submaster $i$ (in group $i$) can compute $\widetilde{\mathbf{A}}_i\vct{x}$ as soon as the task results from any $k_1$ workers within group $i$ are collected.
	Once $\widetilde{\mtx{A}}_{i}\vct{x}$ is computed, it is sent to the master.
	The master can obtain $\mtx{A} \vct{x}$ by retrieving $\widetilde{\mtx{A}}_{i}\vct{x}$ from any $k_2$ submasters.
	For each worker, we define \emph{completion time} as the sum of the runtime of the worker and the time required for delivering its computation result to the submaster.
	For each group, we further define \emph{intra-group latency} as the time for completing its assigned subtask.
	The total computation time is defined as the time from when the workers start to run until the master completes computing $\mtx{A}\vct{x}$. 
	The proposed computation framework can be applied to practical multi-rack systems where
	the input data $\mtx{A}$ is coded and distributed into $n_2$ racks; the $i\textsuperscript{th}$ rack contains $\widetilde{\mtx{A}}_{i}$. 
	For instance, in the Facebook's warehouse cluster, data is encoded with a $(14, 10)$ MDS code, and then the 14 encoded chunks are stored across different racks \cite{Rashmi}. 
	Once $\vct{x}$ is given from the master,  
	the $i\textsuperscript{th}$ rack can compute $\widetilde{\mtx{A}}_{i}\vct{x}$ using the coded data $\widetilde{\mtx{A}}_{i}$ that it contains. 

	\subsection{Application: Matrix-Matrix Multiplications}
Our scheme can be also applied to matrix-matrix multiplications. More specifically, consider computing $\mtx{A}^T\mtx{B}$ for given matrices $\mtx{A}$ and  $\mtx{B} = [\vct{b}_1 \, \vct{b}_2 \, \cdots \, \vct{b}_{k_2}]$. 
After applying an $(n_2, k_2)$ MDS code to $\mtx{B}$, we have $\check{\mtx{B}} = [\check{\vct{b}}_1 \, \check{\vct{b}}_2\, \cdots \, \check{\vct{b}}_{n_2}]$. 
Moreover, group $i$ divides $\mtx{A}$ into $k_1^{(i)}$ equal-sized submatrices as 
$\mtx{A}=[\mtx{A}_{i,1} \, \mtx{A}_{i,2} \, \cdots \, \mtx{A}_{i, k_1^{(i)}}]$, and we apply an $(n_1^{(i)}, k_1^{(i)})$ MDS code, resulting in 
$\check{\mtx{A}}_i =[\check{\mtx{A}}_{i,1}  \, \check{\mtx{A}}_{i,2}  \, \cdots \, \check{\mtx{A}}_{i,n_1^{(i)}} ]$.
The computation $\check{\mtx{A}}_{i,j}^T \check{\vct{b}}_i$ is assigned to worker $w(i, j)$. 
Using an $(n_1^{(i)},k_1^{(i)})$ MDS code, submaster $i$ can compute $\mtx{A}^T \check{\vct{b}}_{i}$ when any $k_1^{(i)}$ workers within its group delivered their computation results. The master can calculate $\mtx{A}^T \mtx{B}$ by gathering $\mtx{A}^T \check{\vct{b}}_{i}$ results from any $k_2$ submasters, using the $(n_2,k_2)$ MDS code.	
Under the homogeneous setting of $n_1^{(i)} = n_1$ and $k_1^{(i)} = k_1$ for all $i \in [n_2]$, the encoding algorithm of the proposed scheme reduces to that of the product coded scheme \cite{Lee17ISIT}. However, the suggested scheme with the homogeneous setting is shown to reduce the decoding cost compared to the product coded scheme under the hierarchical computing structure: a detailed analysis is in Sec.~\ref{sec: deccomp}.

	\section{Latency Analysis} \label{sec: latency}
	\begin{figure}[t]
		\centering
		\includegraphics[width=0.6\linewidth]{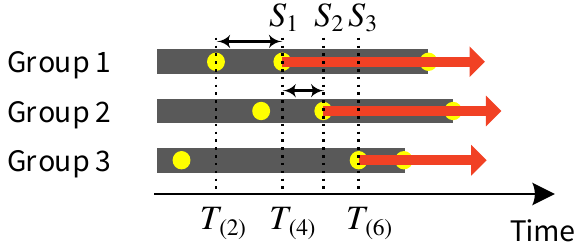}
		\caption{Illustration of obtaining $\mathcal{L}$ in a $(3, 2)\times(3, 2)$ coded computation}
		\vspace{-0.1in}
		\label{fig: lb_ex}
	\end{figure}

We start by providing some preliminaries for the order statistics. 
For $n$ random variables, the $k\textsuperscript{th}$ order statistic is defined by the $k\textsuperscript{th}$ smallest one of $n$.
From the known results from the order statistics~\cite{David03},
the expected value of the $k\textsuperscript{th}$ order statistics out of $n$  $(n>k)$ exponential random variables with rate $\mu$
is
$( H_{n} - H_{n - k})/\mu$,
where $H_{k} = \sum^k_{l = 1} \frac{1}{l} \simeq \log k + \gamma$ as $k$ grows for a fixed constant $\gamma$. This leads to 
$(H_{n} - H_{n - k})/\mu \simeq \frac{1}{\mu}\log \frac{n}{n - k}\,.$
For $n = k$, the expected latency is given by $H_n/\mu \simeq (\log n) / \mu$.
Further, define $H_0 \coloneqq 0$ for ease of exposition.

Consider the hierarchical computing system\footnote{For simplicity of analysis, we only consider the homogeneous setting of $n_1^{(i)} = n_1$ and $k_1^{(i)} = k_1$ for all $i \in [n_2]$.} of Fig. \ref{fig: cluster}.
Assume that for $i \in [n_2]$ and $j \in [n_1]$, the completion time $T_{i,j}$ of worker $w(i,j)$ 
is exponentially distributed with rate $\mu_1$ (i.e., $\Pr[T_{i,j} \leq t]  = 1 - e^{-\mu_1 t}$). 
Further, the communication time $T_i^{(c)}$ from the $i\textsuperscript{th}$ group to the master is also exponentially distributed with rate $\mu_2$ (i.e., $\Pr[T_i^{(c)} \leq t] = 1 - e^{-\mu_2 t}$). Here, we assume that all latencies are independent with one another. Given the assumptions, the total computation time of the $(n_1, k_1) \times (n_2, k_2)$ coded computation is written as
	\begin{equation}\label{eqn:Overall_Latency}
	T = \underset{i \in [n_2]}{k_2\textsuperscript{th} \min} \left(T_i^{(c)}+S_i\right)
	\end{equation}
	where
	\begin{equation}
	S_i = \underset{j \in [n_1]}{k_1\textsuperscript{th} \min } \,\,T_{i,j}
	\end{equation}
	denotes the time to wait for the $k_1$ fastest workers in group $i$.
	The group index $i$ is relabeled such that $S_1 \leq S_2 \leq \cdots\leq S_{n_2}$. In other words, the fastest group that finishes its assigned subtask is relabeled as the $1\textsuperscript{st}$ group, while the slowest group is relabeled as the $n_2\textsuperscript{th}$ group. 
	Here we provide upper and lower bounds on $\mathbb{E}[T]$.


	
	\subsection{Lower Bound}
	
	%
	%
	
	Let $T_{(m)}$ be the $m\textsuperscript{th}$ smallest element of  $\{T_{i,j}\}_{i \in [n_2], j \in [n_1]}$.
	Then, $T_{(1)} \leq T_{(2)} \leq \cdots \leq T_{(n_1 n_2)}$ holds.
	Using this notation, we derive a lower bound on $\mathbb{E}[T]$, formally stated below.
	
	\begin{theorem}\label{thm:lower}{\em
			The expected total computation time of the $(n_1, k_1) \times (n_2, k_2)$ coded computation is lower bounded as
			\begin{equation}\label{eqn: lower}
			\mathbb{E}[T] \geq \mathbb{E} \left[\underset{i \in [n_2]}{k_2\textsuperscript{th} \min} \left(T_i^{(c)}+T_{(i k_1)}\right)\right] \coloneqq \mathcal{L}\,.
			\end{equation}
		} 
	\end{theorem}
	\begin{IEEEproof}
		Consider a realization of $\{T_{i,j}\}_{i \in [n_2], j \in [n_1]}$ and $\{T_i^{(c)}\}_{i\in[n_2]}$. 
		Recall that a group finishes its assigned subtask if $k_1$ workers within the group complete their tasks.	
		Hence, it is impossible for the $i\textsuperscript{th}$ group to finish its work if the total number of completed workers in the system is less than $ik_1$. In other words, it must hold that  
		\begin{equation}\label{eq: TS}
		T_{(ik_1)}  \leq S_i \text{ for all } i\in[n_2]\,.
		\end{equation}
		Thus, the total computation time in (\ref{eqn:Overall_Latency}) should be:
		\begin{equation*}
		T = \underset{i \in [n_2]}{k_2\textsuperscript{th} \min} (T_i^{(c)}+S_i) \geq \underset{i \in [n_2]}{k_2\textsuperscript{th} \min} (T_i^{(c)}+T_{(i k_1)})\,.
		\end{equation*}
		Averaging over all possible realizations, we complete the proof.
	\end{IEEEproof}
	
	\begin{figure}
		\centering
		\resizebox{0.5\textwidth}{!}{%
			\begin{tikzpicture}[->,>=stealth,shorten >=1pt,auto ,node distance=1.5cm, semithick,
			type1cut/.style={black, dashed, semithick, -},
			annot/.style={->, thick},
			]
			\tikzstyle{state}=[circle, fill=white,draw=black,thick,text=black,scale=1, minimum size=0.8cm]
			\tikzstyle{state_red}=[circle, fill={rgb,255: red, 249; green,179; blue,167},draw=black,thick,text=black,scale=1, minimum size=0.8cm]
			\tikzstyle{state_blue}=[circle, fill={rgb,255: red, 149; green,164; blue,222},draw=black,thick,text=black,scale=1, minimum size=0.8cm]
			\tikzstyle{every path}=[font=\footnotesize]
			\node[state_red](S00){$0,0$};
			\node[state](S10)[right of=S00]{$1,0$};
			\node[state](S20)[right of=S10]{$2,0$};
			\node[state](S30)[right of=S20]{$3,0$};
			\node[state](S40)[right of=S30]{$4,0$};
			\node[state](S50)[right of=S40]{$5,0$};
			\node[state](S60)[right of=S50]{$6,0$};
			
			\node[state](S21)[above of=S20]{$2, 1$};
			\node[state](S31)[above of=S30]{$3, 1$};
			\node[state](S41)[above of=S40]{$4, 1$};
			\node[state](S51)[above of=S50]{$5, 1$};
			\node[state](S61)[above of=S60]{$6, 1$};
			
			\node[state_blue](S42)[above of=S41]{$4, 2$};
			\node[state_blue](S52)[above of=S51]{$5, 2$};
			\node[state_blue](S62)[above of=S61]{$6, 2$};
			
			\begin{scope}[on background layer]
			\draw [type1cut] ([yshift=+3.5cm, xshift=0cm] S00 |- S00) -- ([yshift=-0.7cm, xshift=0cm] S00 |- S00);
			\draw [type1cut] ([yshift=+3.5cm, xshift=0cm] S20 |- S20) -- ([yshift=-0.7cm, xshift=0cm] S20 |- S20);
			\draw [type1cut] ([yshift=+3.5cm, xshift=0cm] S40 |- S40) -- ([yshift=-0.7cm, xshift=0cm] S40 |- S40);
			\draw [type1cut] ([yshift=+3.5cm, xshift=0cm] S60 |- S60) -- ([yshift=-0.7cm, xshift=0cm] S60 |- S60);
			\end{scope}
			%
			\node at ([yshift=-0.9cm] S00 |- S00) {$u = 0$};
			\node at ([yshift=-0.9cm] S20 |- S20) {$u = k_1$};
			\node at ([yshift=-0.9cm] S40 |- S40) {$u = k_2k_1$};
			\node at ([yshift=-0.9cm] S60 |- S60) {$u = n_2k_1$};
			
			\node at ([xshift=+1cm] S62 |- S62) {$v = k_2$};
			\node at ([xshift=+1cm] S60 |- S60) {$v = 0$};

			\path (S00) edge [] node[above] {$9\mu_1$} (S10);
			\path (S10) edge [] node[above] {$8\mu_1$} (S20);
			\path (S20) edge [] node[above] {$7\mu_1$} (S30);
			\path (S30) edge [] node[above] {$6\mu_1$} (S40);
			\path (S40) edge [] node[above] {$5\mu_1$} (S50);
			\path (S50) edge [] node[above] {$4\mu_1$} (S60);
			
			\path (S20) edge [] node[left] {$\mu_2$} (S21);
			\path (S30) edge [] node[left] {$\mu_2$} (S31);
			\path (S40) edge [] node[left] {$2\mu_2$} (S41);
			\path (S50) edge [] node[left] {$2\mu_2$} (S51);
			\path (S60) edge [] node[left] {$3\mu_2$} (S61);
			
			\path (S21) edge [] node[above] {$7\mu_1$} (S31);
			\path (S31) edge [] node[above] {$6\mu_1$} (S41);
			\path (S41) edge [] node[above] {$5\mu_1$} (S51);
			\path (S51) edge [] node[above] {$4\mu_1$} (S61);
			
			\path (S41) edge [] node[left] {$\mu_2$} (S42);
			\path (S51) edge [] node[left] {$\mu_2$} (S52);
			\path (S61) edge [] node[left] {$2\mu_2$} (S62);
			\end{tikzpicture}
		}%
		\caption{State transition diagram producing the lower bound $\mathcal{L}$ on the expected latency in a $(3, 2)\times(3,2)$ coded computation. Each state is labeled with $(u,v)$, where $u$ is the number of completed workers and $v$ is the number of groups that have sent their computation results.}
		\vspace{-0.1in}
		\label{fig: MC}
	\end{figure}
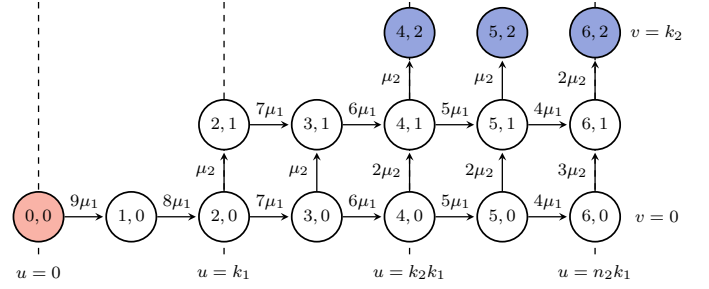
	
	To further illustrate the proof, we provide a schematic example in Fig. \ref{fig: lb_ex}. Consider a $(3,2)\times (3,2)$ coded computation. The yellow circles denote the completion times of the workers. 
	After $k_1 = 2$ workers in a group finish their computations, the group-master communication, shown as the red arrows, starts from each group. As can be seen, $T_{(2)} \leq S_1$, $T_{(4)} \leq S_2$ and $T_{(6)} \leq S_3$, which concur with \eqref{eq: TS}.	
	The following lemma shows that $\mathcal{L}$ can be computed by analyzing the hitting time of an auxiliary Markov chain.
	
	\begin{lemma}\label{lem:markov}
		Let $\mathbb{C}$ be the continuous-time Markov chain defined over the state space $(u,v) \in \{0, 1, \dots, n_2k_1\} \times \{0, 1, \dots, k_2\}$. The state transition rates of $\mathbb{C}$ are defined as follows:
		\begin{itemize}
			\item From state $(u,v)$ to state $(u+1,v)$ at rate $(n_1n_2 - i)\mu_1$,
			if $vk_1 \leq u < n_2k_1$,
			\item From state $(u,v)$ to state $(u,v + 1)$ at rate $\left(\floor*{{u}/{k_1}} - v\right)\mu_2$,
			if $0 \leq v < \min\left\{\floor*{{u}/{k_1}}, k_2\right\}$.
		\end{itemize}
	\end{lemma}
	Then, the expected hitting time of $\mathbb{C}$ from state $(0,0)$ to the set of states $\{(u,k_2)\}_{u=k_2k_1}^{n_2k_1}$ is equal to $\mathcal{L}$.
	
	\begin{IEEEproof}
		See Appendix \ref{proof: MC} for the proof.	
	\end{IEEEproof}

	Markov chain $\mathbb{C}$ defined in Lemma \ref{lem:markov} consists of the states $(u,v)$, where $u$ represents the number of completed workers and $v$ indicates the number of groups which have delivered their computation results to the master.
	
	For an illustrative example, the state transition diagram for a $(3, 2)\times(3, 2)$ coded computation yielding a lower bound is shown in Fig. \ref{fig: MC}.
	The overall computation is terminated when the $k_2 = 2$ groups finish conveying their computational results to the master,
	i.e., when the Markov chain visits the states with $v = 2$ for the first time.
	We see that $u$ increases by one when a worker completes its computation,
	and 
	$v$ increases by one 
	when master receives the computation result from a group.
	The rightward transition (to increase $u$) rate is determined by the product of $\mu_1$ and the number of remaining workers.
	The upward transition (to increase $v$) rate is the product of $\mu_2$ and the number of groups that have not delivered their computation results to the master.
	The proposed lower bound $\mathcal{L}$ can be easily computed from the first-step analysis \cite{Bremaud13} of the Markov chain produced by Lemma \ref{lem:markov}.
	
	\subsection{Upper Bound}
	
	We here provide two upper bounds on the expected total computation time. 
	The first bound in the following lemma is applicable for all values of $n_1$ and $k_1$.
	
	\begin{lemma}
		{\em
			The expected total computation time of the $(n_1, k_1)\times(n_2, k_2)$ coded computation is upper bounded as
			$\mathbb{E}[T] \leq {H_{n_1n_2}}/{\mu_1} + ({H_{n_2} - H_{n_2 - k_2}})/{\mu_2} \,.$
		}\label{lem: n1k1}
	\end{lemma}
	\begin{IEEEproof}
		See Appendix \ref{sec:proofn1k1} for the proof.
	\end{IEEEproof}
	
	
	We now establish another upper bound using the following two steps.
	First we find an upper bound on the maximum intra-group latency among $n_2$ groups.
	Afterwards, adding this value to the expected latency of the group-master communication yields an upper bound on the expected total computation time.    
	For given $n_1$ and $k_1$, we use $\delta_1 > 0$ which satisfies
	$n_1 = (1 + \delta_1)k_1$.
	We now present the asymptotic upper bound as follows.
	
	\begin{theorem}\label{thm: ub}\em{
			For a fixed constant $\delta_1 > 0$, the expected total computation time of the $(n_1, k_1)\times(n_2, k_2)$ coded computing system is upper bounded as
			$\mathbb{E}[T] \leq [\log({1 + \delta_1})/{\delta_1}]/{\mu_1} + ({H_{n_2} - H_{n_2 - k_2}})/{\mu_2} + o(1)$
			in the limit of $k_1$.}	
	\end{theorem}
	\begin{IEEEproof}
		See Appendix \ref{sec:proofub} for the proof.
	\end{IEEEproof}

\subsection{Evaluation of Bounds}
\begin{figure}[t]
	\centering
	\subfloat[\label{fig: smallk1}$n_1 = 10$, $k_1 = 5$, $n_2 = 10$]{
		\includegraphics[width=0.5\linewidth]{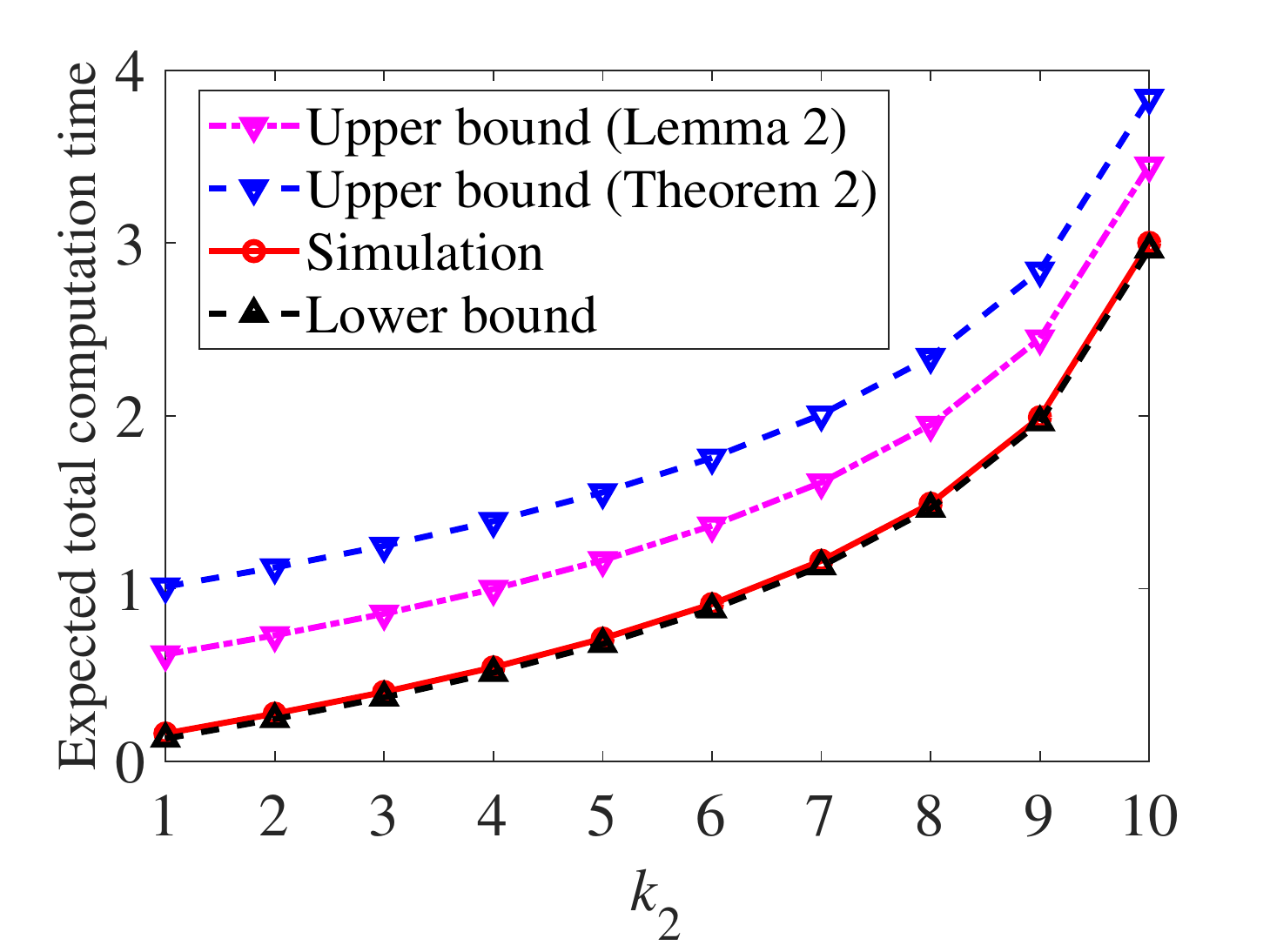}}
	\subfloat[\label{fig: largek1}$n_1 = 600$, $k_1 = 300$, $n_2 = 10$]{
		\includegraphics[width=0.5\linewidth]{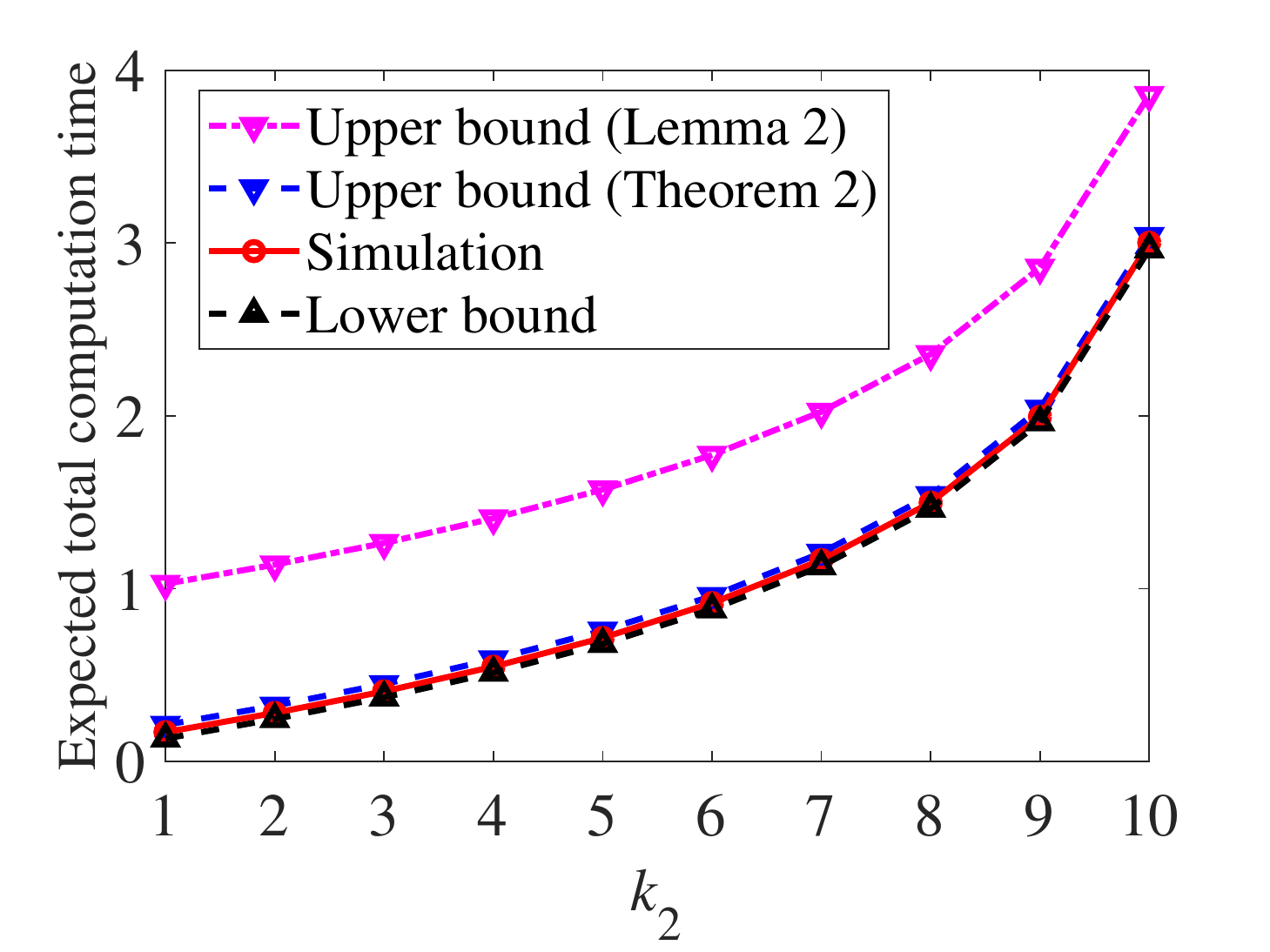}}
	\caption{The expected total computation time of the $(n_1, k_1) \times (n_2,k_2)$ coded computing with its bounds for varying $k_2$}
	\label{fig: boundsk2}
\end{figure}
Fig. \ref{fig: boundsk2} shows the behavior of the expected total computation time and its upper/lower bounds with varying $k_2$.
Here we consider two upper bounds proposed in Lemma \ref{lem: n1k1} and Theorem \ref{thm: ub}.
To see the impact of $k_1$, the values of $k_1$ are fixed to 5 and 300 in Figs. \ref{fig: smallk1} and \ref{fig: largek1}, respectively.
The other code parameters are set to $n_1 = (1 + \delta_1)k_1, n_2 = 10$ for both figures, where $\delta_1$ is fixed to $1$.
The rates of the completion time of the worker and group-master communication are set to $\mu_1 = 10$ and $\mu_2 = 1$.
For a relatively small values of $k_1$, the upper bound in Lemma \ref{lem: n1k1} is a tighter upper bound than the upper bound in Theorem \ref{thm: ub}.
As can be seen in Fig. \ref{fig: largek1}, the asymptotic upper bound in Theorem \ref{thm: ub} becomes tighter as $k_1$ grows, which
concurs with Theorem \ref{thm: ub}.  
We also have numerically confirmed that the proposed lower bound is tight.

	\section{Decoding Complexity}\label{sec: deccomp}
	In this section, we compare decoding complexity of our hierarchical coding with the replication and non-hierarchical coding schemes including the $(n_1,k_1)\times(n_2,k_2)$ product code \cite{Lee17ISIT} and the $(n,k)$ polynomial code \cite{Yu17}.
	For fair comparison, we set $n = n_1n_2$ and $k = k_1k_2$.
	We further assume that the decoding complexity of the $(n,k)$ MDS code is $\mathcal{O}(k^\beta)$ for some $\beta > 1$.\footnote{Note that this is the case for most practical decoding algorithms~\cite{Halbawi16ISIT, Halbawi16ITW}. Decoding with $\beta = 1$ requires a large field size~\cite{Yu17}.}
	In our framework, the $n_2$ intra-group codes can be decoded in parallel, followed by decoding of the cross-group code using the $k_2$ fastest results.
	Thus, the overall decoding procedure consists of 1) parallel decoding of $(n_1, k_1)$ intra-group MDS codes and 2) decoding of the $(n_2, k_2)$ cross-group MDS code, resulting in the total decoding cost of $\mathcal{O}(k_1^\beta + k_1k_2^\beta)$.
	Similarly, one can show that the decoding cost of polynomial codes is $\mathcal{O}(k^\beta)$, and that of product code is $\mathcal{O}(k_1k_2^\beta + k_2k_1^\beta)$. 
We note that the hierarchical code can have a substantial improvement, sometimes by an order of magnitude, in decoding complexity, compared to the product code. 
For instance, if $\beta = 2$ and $k_1 = k_2^2$, the decoding cost of hierarchical code becomes $\mathcal{O}(k_2^{4})$ while that of the product code is $\mathcal{O}(k_2^{5})$; if $k_1 = k_2^{1.5}$, the decoding costs are $\mathcal{O}(k_2^{3.5})$ and $\mathcal{O}(k_2^{4})$, respectively.
In general, if $k_1 = k_2^p$, one can show that the relative gain of the hierarchical codes in decoding cost monotonically increases as $p$ increases, providing a guideline for efficient code designs.
%
%
Table \ref{table: total_compare} summarizes the computing times and decoding costs of various coding schemes.

\begin{figure}[t]
	\centering
	\includegraphics[width=0.6\linewidth]{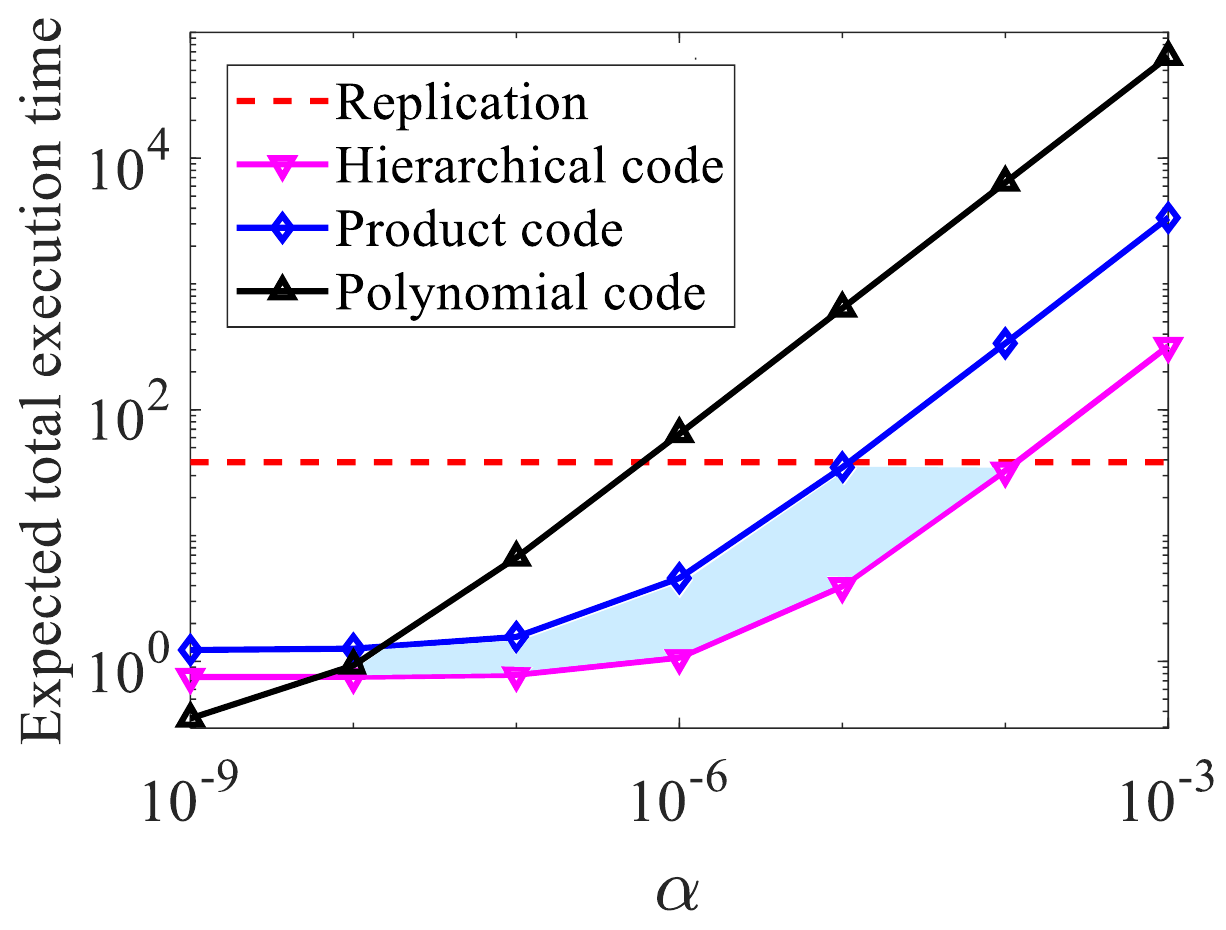}
	\vspace{-0.05in}
	\caption{$\mathbb{E}[T_\text{exec}]$ of various coding schemes for parameters of $(n_1, k_1) = (800,400)$, $(n_2,k_2)=(40,20), (\mu_1,\mu_2) = (10,1)$ and $\beta=2$}
	\vspace{-0.15in}
	\label{fig: dec}
\end{figure}

We now compare the expected total execution time defined as $T_\text{exec} \coloneqq T_\text{comp}+ \alpha T_\text{dec}$, where $T_\text{comp}$ is the computing time, $T_\text{dec}$ is the decoding cost, and $\alpha \geq 0$ is the relative weight of the decoding cost.
We note that $\alpha$ is a system-specific parameter that depends on 1) the relative CPU speed of the master compared to the workers and 2) dimension of the input data.
Shown in Fig. \ref{fig: dec} are the expected total execution times for parameters of $(n_1, k_1) = (800,400)$, $(n_2,k_2)=(40,20), (\mu_1,\mu_2) = (10,1)$ and $\beta=2$.

We first observe that with all tested practical values of $\mu_1$ and $\mu_2$, the hierarchical code strictly outperforms the product code for all values of $\alpha$. 
Further, we observe that the optimal choice of coding scheme depends on the value of $\alpha$ as follows:
\begin{itemize}
\item (moderate $\alpha$) when both $T_\text{comp}$ and $T_\text{dec}$ have to be minimized, the hierarchical code achieves the lowest $T_\text{exec}$ by striking a balance between them;
\item (low $\alpha$) when $T_\text{dec}$ is negligible, the polynomial code achieves the lowest $T_\text{exec}$; and
\item (high $\alpha$) when $T_\text{dec}$ dominates $T_\text{exec}$, the replication code is the best.
\end{itemize}
Note that the shaded area in Fig. \ref{fig: dec} represents the additional achievable $(\alpha, \mathbb{E}[T_\text{exec}])$ region thanks to introducing the hierarchical code.

	\begin{table}[t]
		\vspace{-0.15in}
		\caption{Comparisons of various coding schemes}
		\vspace{-0.2in}
		\begin{center}
			\begin{tabular} {ccc}
				\hline\hline
				Coding scheme & Computing time  ($T_\text{comp}$)  &  Decoding cost ($T_\text{dec}$)   \\
				\hline
				 Replication&  $kH_{k}/(n\mu_2)$ &0 \\
				 Hierarchical code & $\mathbb{E}[T]$ &  $\mathcal{O}(k_1k_2^\beta + k_1^\beta)$ \\
				 Product code \cite{Lee17ISIT} & $\frac{1}{\mu_2} \log\left(\frac{\sqrt{n/k} + \sqrt[\leftroot{-2}\uproot{2}4]{n/k}}{\sqrt{n/k} - 1}\right)$  & $\mathcal{O}(k_1k_2^\beta + k_2k_1^\beta)$ \\
				 Polynomial code \cite{Yu17} & $(H_n - H_{n-k})/\mu_2$& $\mathcal{O}(k_1^\beta k_2^\beta)$\\
				\hline\hline
			\end{tabular}
		\vspace{-0.15in}
		\end{center}\label{table: total_compare}
	\end{table}



	 \appendices
	 \section{Proof of Lemma \ref{lem:markov}}\label{proof: MC} 
	 
\begin{figure}[t]
	 	\centering
	 	\includegraphics[width=0.9\linewidth]{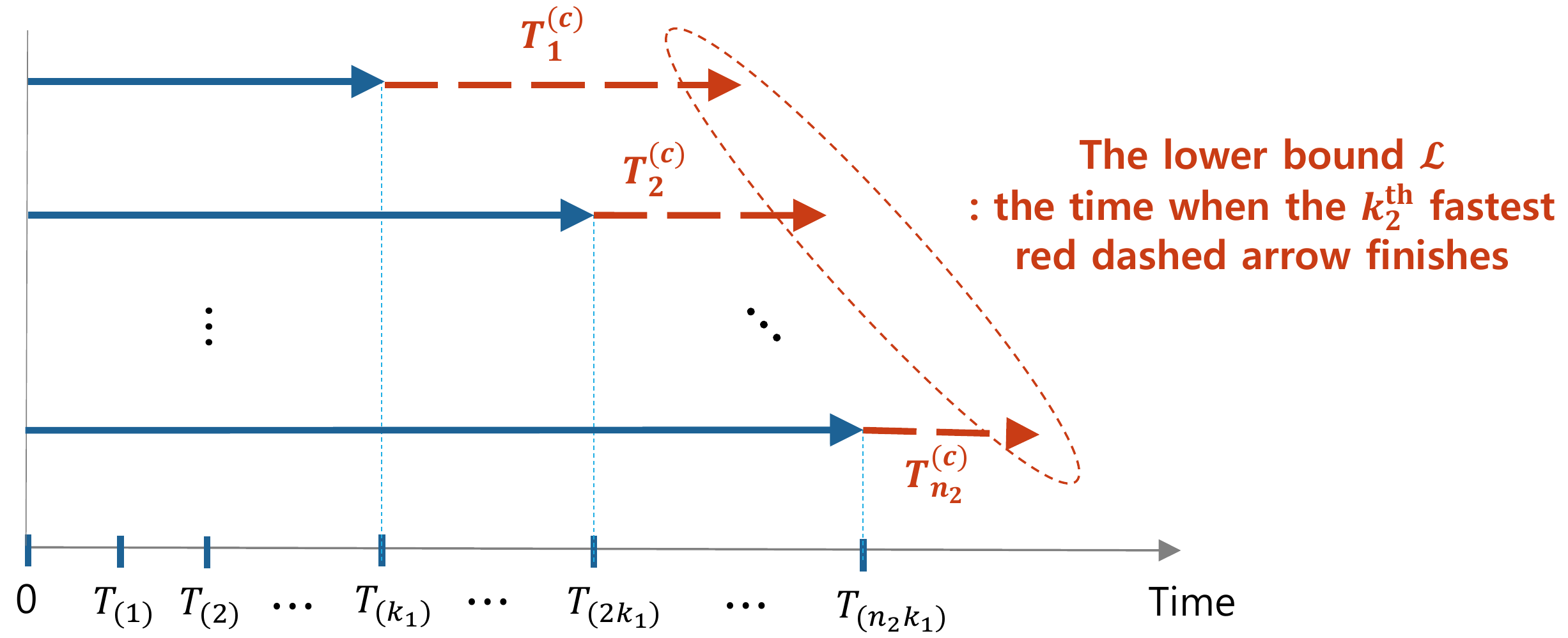}
	 	\caption{Illustration of the lower bound $\mathcal{L}$}
	 	\label{fig: prop1}
	 	\vspace{-0.1in}
	 \end{figure}

	 Note that the lower bound $\mathcal{L}$ in Theorem \ref{thm:lower} can be illustrated as in Fig. \ref{fig: prop1}. The lower bound depends on two types of variables: $\{T_{(m)}\}_{m=1}^{n_2 k_1}$, the set of $n_2k_1$ smallest realizations of $n_2n_1$ exponentially distributed random variables with rate $\mu_1$ and $\{T_c^{(l)}\}_{l=1}^{n_2}$, the set of $n_2$ exponentially distributed random variables with rate $\mu_2$.
	 Consider arbitrary realizations of $\{T_{(m)}\}$ and $\{T_c^{(l)}\}$. 	
	 For a given time $t$, define
	 \begin{align}
	 u &\coloneqq \max_{m \in [n_2k_1]} \{t \geq T_{(m)}\}, \label{eqn: mcrow}\\
	 v &\coloneqq \left| \{l \in [n_2] : t \geq T_c^{(l)} + T_{(lk_1)} \}  \right| \label{eqn: mccol}.
	 \end{align}
	 Thus, each time slot $t$ can be assigned to a state $(u,v)$ for $u \in \{0, 1, \cdots, n_2k_1\}$ and $v \in \{0, 1, \cdots, n_2\}$.
	 From the definition of $\mathcal{L}$ in (\ref{eqn: lower}), the lower bound corresponds to the expected time $t$ to achieve $v = k_2$. Thus, we consider the state space of $(u,v) \in \{0, 1, \cdots, n_2k_1\} \times \{0, 1, \cdots, k_2\}$, and find the expected time to arrive at states $(u,v)$ with $v=k_2$ from state $(0,0)$.
	 
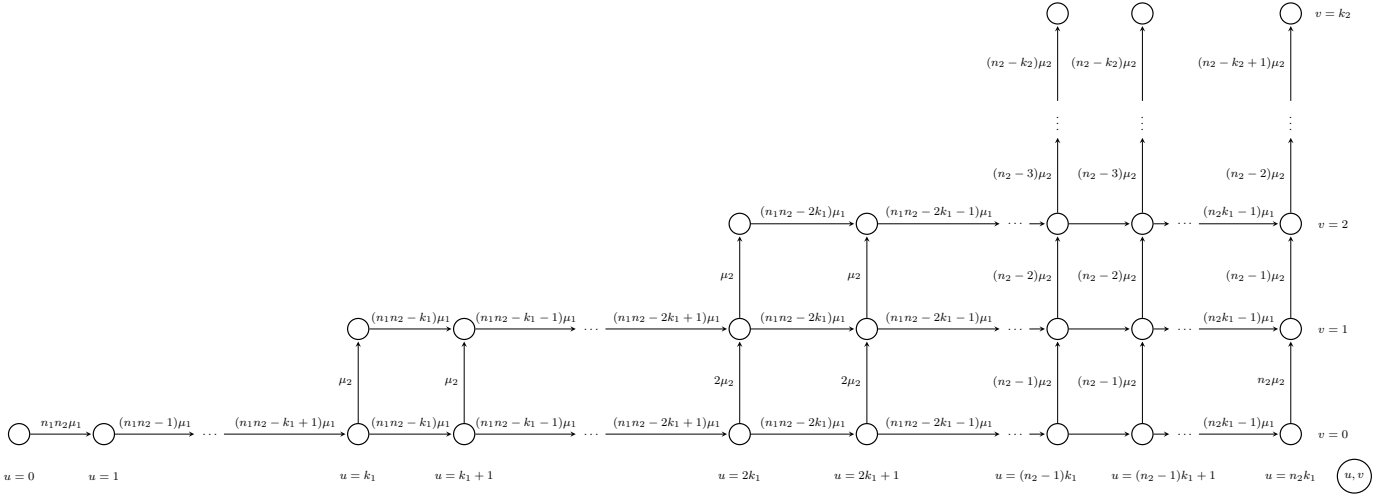
\begin{figure*}[t]
	 	\centering
	 	\resizebox{1\textwidth}{!}{%
	 		\begin{tikzpicture}[->,>=stealth,shorten >=1pt,auto ,node distance=3.5cm, semithick,
	 		type1cut/.style={black, dashed, semithick, -},
	 		annot/.style={->, thick},
	 		]
	 		\tikzstyle{state}=[circle,fill=white,draw=black,thick,text=black,scale=1, minimum size=0.5cm]
	 		\tikzstyle{state2}=[circle,fill=white,draw=white,thick,text=black,scale=1, minimum size=0.5cm]
	 		\tikzstyle{every path}=[font=\footnotesize]
	 		\node[state](S00){};
	 		\node[state](S10)[xshift = -1.5cm, right of=S00]{};
	 		\node[state2](Sdots00)[xshift = -1cm, right of=S10]{$\ldots$};
	 		\node[state](S20)[right of=Sdots00]{};
	 		\node[state](S30)[xshift = -1cm, right of=S20]{};
	 		\node[state2](Sdots10)[xshift = -0.5cm, right of=S30]{$\ldots$};
	 		\node[state](S40)[right of=Sdots10]{};
	 		\node[state](S50)[xshift = -0.5cm, right of=S40]{};
	 		\node[state2](Sdots20)[right of=S50]{$\ldots$};
	 		\node[state](S60)[xshift = -2.5cm, right of=Sdots20]{};
	 		\node[state](S70)[xshift = -1.5cm, right of=S60]{};
	 		\node[state2](Sdots30)[xshift = -2.5cm, right of=S70]{$\ldots$};
	 		\node[state](S80)[xshift = -1cm, right of=Sdots30]{};
	 		
	 		\node[state](S21)[yshift = -1cm, above of=S20]{};
	 		\node[state](S31)[yshift = -1cm, above of=S30]{};
	 		\node[state2](Sdots11)[yshift = -1cm, above of=Sdots10]{$\ldots$};
	 		\node[state](S41)[yshift = -1cm, above of=S40]{};
	 		\node[state](S51)[yshift = -1cm, above of=S50]{};
	 		\node[state2](Sdots21)[yshift = -1cm, above of=Sdots20]{$\ldots$};
	 		\node[state](S61)[yshift = -1cm, above of=S60]{};
	 		\node[state](S71)[yshift = -1cm, above of=S70]{};
	 		\node[state2](Sdots31)[yshift = -1cm, above of=Sdots30]{$\ldots$};
	 		\node[state](S81)[yshift = -1cm, above of=S80]{};
	 		
	 		\node[state](S42)[yshift = -1cm, above of=S41]{};
	 		\node[state](S52)[yshift = -1cm, above of=S51]{};
	 		\node[state2](Sdots22)[yshift = -1cm, above of=Sdots21]{$\ldots$};
	 		\node[state](S62)[yshift = -1cm, above of=S61]{};
	 		\node[state](S72)[yshift = -1cm, above of=S71]{};
	 		\node[state2](Sdots32)[yshift = -1cm, above of=Sdots31]{$\ldots$};
	 		\node[state](S82)[yshift = -1cm, above of=S81]{};
	 		
	 		\node[state2](Sdots33)[yshift = -1cm, above of=S62]{$\vdots$};
	 		\node[state2](Sdots43)[yshift = -1cm, above of=S72]{$\vdots$};
	 		\node[state2](Sdots53)[yshift = -1cm, above of=S82]{$\vdots$};
	 		
	 		\node[state](S63)[yshift = -1cm, above of=Sdots33]{};
	 		\node[state](S73)[yshift = -1cm, above of=Sdots43]{};
	 		\node[state](S83)[yshift = -1cm, above of=Sdots53]{};
	 		
	 		
	 		%
	 		\node at ([yshift=-1cm] S00 |- S00) {$u = 0$};
	 		\node at ([yshift=-1cm] S10 |- S10) {$u = 1$};
	 		\node at ([yshift=-1cm] S20 |- S20) {$u = k_1$};
	 		\node at ([yshift=-1cm] S30 |- S30) {$u = k_1 + 1$};
	 		\node at ([yshift=-1cm] S40 |- S40) {$u = 2k_1$};
	 		\node at ([yshift=-1cm] S50 |- S50) {$u = 2k_1 + 1$};
	 		\node at ([yshift=-1cm, xshift = -0.5cm] S60 |- S60) {$u = (n_2 - 1)k_1$};
	 		\node at ([yshift=-1cm, xshift = +0.5cm] S70 |- S70) {$u = (n_2 - 1)k_1 + 1$};
	 		\node at ([yshift=-1cm] S80 |- S80) {$u = n_2k_1$};
	 		
	 		\node at ([xshift=1cm] S80 |- S80) {$v = 0$};
	 		\node at ([xshift=1cm] S81 |- S81) {$v = 1$};
	 		\node at ([xshift=1cm] S82 |- S82) {$v = 2$};
	 		\node at ([xshift=1cm] S83 |- S83) {$v = k_2$};
	 		
	 		\node[state](Sij)[xshift= -2cm, yshift= -1cm, right of=S80]{$u, v$};
	 		
	 		\path (S00) edge [] node[above] {$n_1n_2\mu_1$} (S10);
	 		\path (S10) edge [] node[above] {$(n_1n_2 - 1)\mu_1$} (Sdots00);
	 		\path (Sdots00) edge [] node[above] {$(n_1n_2 - k_1+1)\mu_1$} (S20);
	 		\path (S20) edge [] node[above] {$(n_1n_2 - k_1)\mu_1$} (S30);
	 		\path (S30) edge [] node[above] {$(n_1n_2 - k_1-1)\mu_1$} (Sdots10);
	 		\path (Sdots10) edge [] node[above] {$(n_1n_2 - 2k_1+1)\mu_1$} (S40);
	 		\path (S40) edge [] node[above] {$(n_1n_2 - 2k_1)\mu_1$} (S50);
	 		\path (S50) edge [] node[above] {$(n_1n_2 - 2k_1-1)\mu_1$} (Sdots20);
	 		\path (Sdots20) edge [] node[above] {} (S60);
	 		\path (S60) edge [] node[above] {} (S70);
	 		\path (S70) edge [] node[above] {} (Sdots30);
	 		\path (Sdots30) edge [] node[above] {$(n_2k_1 - 1)\mu_1$} (S80);

	 		\path (S21) edge [] node[above] {$(n_1n_2 - k_1)\mu_1$} (S31);
	 		\path (S31) edge [] node[above] {$(n_1n_2 - k_1 - 1)\mu_1$} (Sdots11);
	 		\path (Sdots11) edge [] node[above] {$(n_1n_2-2k_1 + 1)\mu_1$} (S41);
	 		\path (S41) edge [] node[above] {$(n_1n_2 - 2k_1)\mu_1$} (S51);
	 		\path (S51) edge [] node[above] {$(n_1n_2 - 2k_1-1)\mu_1$} (Sdots21);
	 		\path (Sdots21) edge [] node[above] {} (S61);
	 		\path (S61) edge [] node[above] {} (S71);
	 		\path (S71) edge [] node[above] {} (Sdots31);
	 		\path (Sdots31) edge [] node[above] {$(n_2k_1 - 1)\mu_1$} (S81);
	 		
	 		\path (S42) edge [] node[above] {$(n_1n_2 - 2k_1)\mu_1$} (S52);
	 		\path (S52) edge [] node[above] {$(n_1n_2 - 2k_1-1)\mu_1$} (Sdots22);
	 		\path (Sdots22) edge [] node[above] {} (S62);
	 		\path (S62) edge [] node[above] {} (S72);
	 		\path (S72) edge [] node[above] {} (Sdots32);
	 		\path (Sdots32) edge [] node[above] {$(n_2k_1 - 1)\mu_1$} (S82);

	 		\path (S20) edge [] node[left] {$\mu_2$} (S21);
	 		\path (S30) edge [] node[left] {$\mu_2$} (S31);
	 		\path (S40) edge [] node[left] {$2\mu_2$} (S41);
	 		\path (S50) edge [] node[left] {$2\mu_2$} (S51);
	 		\path (S60) edge [] node[left] {$(n_2 - 1)\mu_2$} (S61);
	 		\path (S70) edge [] node[left] {$(n_2 - 1)\mu_2$} (S71);
	 		\path (S80) edge [] node[left] {$n_2\mu_2$} (S81);
	 		
	 		\path (S41) edge [] node[left] {$\mu_2$} (S42);
	 		\path (S51) edge [] node[left] {$\mu_2$} (S52);
	 		\path (S61) edge [] node[left] {$(n_2 - 2)\mu_2$} (S62);
	 		\path (S71) edge [] node[left] {$(n_2 - 2)\mu_2$} (S72);
	 		\path (S81) edge [] node[left] {$(n_2-1)\mu_2$} (S82);
	 		
	 		\path (S62) edge [] node[left] {$(n_2 - 3)\mu_2$} (Sdots33);
	 		\path (S72) edge [] node[left] {$(n_2 - 3)\mu_2$} (Sdots43);
	 		\path (S82) edge [] node[left] {$(n_2-2)\mu_2$} (Sdots53);
	 		
	 		\path (Sdots33) edge [] node[left] {$(n_2 - k_2)\mu_2$} (S63);
	 		\path (Sdots43) edge [] node[left] {$(n_2 - k_2)\mu_2$} (S73);
	 		\path (Sdots53) edge [] node[left] {$(n_2-k_2 + 1)\mu_2$} (S83);

	 		\end{tikzpicture}
	 	}%
	 	\caption{State transition diagram for the $(n_1, k_1)\times(n_2, k_2)$ coded computation producing a lower bound. Any state is denoted by $(u, v)$, where $u$ describes the number of completed workers and $v$ is the number of groups that sent their computation results.}
	 	\label{fig: MCfull}

	 \end{figure*}

	 We now examine the state transition rates.
	 From the definitions of $T_{(m)}$ and (\ref{eqn: mcrow}), the transition from state $(u,v)$ to state $(u+1,v)$ occurs with rate $(n_1 n_2-u)\mu_1$, since there are $n_1 n_2 - u$ remaining $\{T_{(m)}\}_{m=u+1}^{n_1n_2}$ such that $t < T_{(m)}$ holds.
	 Moreover, for a given time $t$ and the corresponding state $(u,v)$, we have $\{T_{(lk_1)}\}_{l=1}^{\floor*{u/k_1}}$ which satisfies $t \geq T_{(lk_1)}$, and $v$ in (\ref{eqn: mccol}) is expressed as 
	 \begin{equation}
	 v = \left| \{l \in \{1, 2, \cdots, \floor*{u/k_1} \} : t \geq T_c^{(l)} + T_{(lk_1)} \}  \right|
	 \end{equation}
	 since $T_c^{(l)}$ is a random variable with nonnegative values.
	 Thus, out of $\floor*{u/k_1}$ activated (i.e., $t \geq T_{(lk_1)}$) random variables $\{T_{(lk_1)}\}_{l=1}^{\floor*{u/k_1}}$, only $v$ random variables satisfy $t \geq T_c^{(l)} + T_{(lk_1)}$. 
	 Therefore, the transition from state $(u, v)$ to state $(u, v + 1)$ occurs with rate $(\floor*{u/k_1} - v)\mu_2$, 
	 for $v \in \{0, 1, \cdots, \min\{\floor*{u/k_1}, k_2\} - 1\}$. Fig. \ref{fig: MCfull} shows the consequent state transition diagram. This Markov chain is identical to $\mathbb{C}$, which completes the proof. 
	 
 \section{Proof of Lemma \ref{lem: n1k1}}\label{sec:proofn1k1}
 $H_{n_1n_2}/\mu_1$ is the maximum intra-group latency, which comes from waiting for all $n_1n_2$ workers. 
 Assuming that every group starts the group-master communication at time $H_{n_1n_2}/\mu_1$, the expected total computation time can be obtained by summing up the group-master communication time to $H_{n_1n_2}/\mu_1$. 
 The group-master communication time is calculated from the time that the $k_2\textsuperscript{th}$ fastest group finishes communication to the master, which is given by $(H_{n_2} - H_{n_2 - k_2})/\mu_2$.
 This completes the proof.  
 
 \section{Proof of Theorem \ref{thm: ub}}\label{sec:proofub}

 A part of the proof generalizes the idea of \cite{Lee17ISIT}, which analyzes the latency of the product code.
 First, we focus on the intra-group latency of each group.
 The expected latency of the $k_1\textsuperscript{th}$ fastest worker out of $n_1$ is given by
 $( H_{n_1} - H_{n_1 - k_1})/\mu_1$,
 where the latency of a worker assumes an exponential distribution with rate $\mu_1$.
 %
 Noticing that the expected completion time of a worker $( H_{n_1} - H_{n_1 - k_1})/\mu_1$ is rewritten as $\frac{1}{\mu_1} \log \frac{1 + \delta_1}{\delta_1}$ for a fixed constant $\delta_1 > 0$ and a sufficiently large $n_1$, 
 we define $$t_0 := \frac{1}{\mu_1} \log \frac{1 + \delta_1}{\delta_1} + \alpha \sqrt{\frac{\log k_1}{k_1}}$$ for some constant $\alpha > 0$.
 
 Consider group $i_0$ with $n_1$ workers.
 Then, for worker $w(i_0, j)$, assume a Bernoulli random variable $X_{i_0, j}$ which takes $0$ when worker $w(i_0,j)$ has completed its computation by time $t_0$, and takes $1$ otherwise.
 Then, probability $p_0$ that $X_{i_0, j}$ takes 1 is:
 \begin{align}
 p_0 \coloneqq \Pr[T_{i_0, j} > t_0] &= e^{-\mu_1t_0}= \frac{\delta_1}{1 + \delta_1} e^{- \mu_1 \alpha \sqrt{\frac{\log k_1}{k_1}}}\nonumber\\
 &\simeq \frac{\delta_1}{1 + \delta_1} \left( 1 - \mu_1\alpha\sqrt{\frac{\log k_1}{k_1}} \right)\label{eq: appre} \,,
 \end{align}
 where \eqref{eq: appre} follows because $\mu_1\alpha\sqrt{\frac{\log k_1}{k_1}}$ is quite small with a sufficiently large $k_1$.	Out of $n_1$ workers in group $i_0$, the set of workers not completed by time $t_0$ is represented as
 \begin{equation*}
 S_{t_0} = \{j \in [n_1] :  T_{i_0, j} > t_0\}= \{ j \in [n_1] : X_{i_0, j} = 1\}
 \end{equation*} 
 with $ \lvert S_{t_0} \rvert$ representing the number of workers not completed by time $t_0$, where $\lvert \cdot \rvert$ denotes the cardinality of a set.

 Since $X_{i_0, j}$ is a Bernoulli random variable with parameter $p_0$,
 the expected number of workers not completed in group $i_0$ is calculated as 
 $n_1p_0 = \delta_1(k_1 - \mu_1 \alpha \sqrt{k_1 \log k_1})$
 for a given $n_1 = (1 + \delta_1)k_1$.
 Recall that a group finishes its assigned subtask when $k_1$ out of 
 $n_1$ workers in the group completed their works. At time $t_0$, we thus denote a case where the number of stragglers in group $i$ is greater than $\delta_1k_1 = n_1 - k_1$ by an error event $E_i$ for $i \in [n_2]$.
 For group $i_0$, we wish to find an upper bound on the probability that $E_{i_0}$ occurs, which is equivalent to the probability that group $i_0$ has not finished its assigned subtask by time $t_0$.  
 We establish such a bound using Hoeffding's inequality \cite{Hoeffding63}
 to bound the deviation of $\lvert S_{t_0} \rvert$ from the mean:
 $$\Pr[\lvert S_{t_0} \rvert - \delta_1(k_1 - \mu_1 \alpha \sqrt{k_1 \log k_1}) \geq t ] \leq e^{ -\frac{2t^2}{(1 + \delta_1)k_1}}\,.$$
 By setting $t = \delta_1\mu_1\alpha\sqrt{k_1 \log k_1}$,
 we obtain
 \begin{align}
 \Pr\left[\lvert S_{t_0} \rvert \geq \delta_1 k_1 \right] &\leq e^{ -\frac{2\delta_1^2\mu_1^2\alpha^2}{1 + \delta_1} \log k_1 } = k_1^{-\frac{2\delta_1^2\mu_1^2\alpha^2}{1 + \delta_1}}\,.\nonumber
 \end{align}
 Combining all $n_2$ groups,   
 the upper bound on the probability that $n_2$ groups not finished their assigned subtasks by time $t_0$ is obtained by the union bound.
 Let $T_I$ be the time when all $n_2$ groups finish their assigned subtasks. Hence, we have
 \begin{align}
 \Pr[T_I > t_0] &=\Pr[E_1\cup E_2\cup \cdots \cup E_{n_2}]\\
 &\leq \sum^{n_2}_{i = 1} \Pr[E_i] = n_2 k_1^{-\frac{2\delta_1^2\mu_1^2\alpha^2}{1 + \delta_1}}= o(k_1^{-1})\label{eq: o}\,,
 \end{align}
 where the last equality holds since $\alpha$ can be made arbitrarily large. 
 Then the expected intra-group latency $\mathbb{E}[T_I]$ satisfies
 \begin{align}
 \mathbb{E}[T_I] &\leq \Pr[T_I \leq t_0] t_0 + \Pr[T_I > t_0] \left(\frac{H_{n_1n_2}}{\mu_1} + t_0\right) \label{eq: max}\\
 & = \left(1 - o(k_1^{-1})\right)t_0 + o(k_1^{-1}) \left(\frac{H_{n_1n_2}}{\mu_1} + t_0\right)\\
 & = \frac{1}{\mu_1} \log \frac{1 + \delta_1}{\delta_1} + o(1) \,, \label{eq: T1}
 \end{align}
 where \eqref{eq: max} is due to the fact that
 $t_0$ is the worst case latency for all events satisfying $T_I \leq t_0$, and 
 $H_{n_1n_2}/\mu_1 + t_0$ is an upper bound on $\mathbb{E}[T_I|T_I>t_0]$.
 
 From \eqref{eq: T1}, we conclude that all the $n_2$ groups embark on the group-master communication before time $\frac{1}{\mu_1} \log \frac{1 + \delta_1}{\delta_1}$, as $k_1$ grows large.
 Hence, adding
 the latency of the group-master communication 
 $(H_{n_2} - H_{n_2 - k_2})/{\mu_2}$ 
 to \eqref{eq: T1} gives an upper bound on the expected total computation time.
 This completes the proof of the case where $n_2 > k_2$. 
 When $n_2 = k_2$, the group-master communication time is represented by ${H_{n_2}}/{\mu_2}$.
 Thus, adding this value to (\ref{eq: T1}) completes the proof, using $H_0 = 0$.

	\bibliographystyle{IEEEtran}
	\bibliography{clusterComputationISIT}
\end{document}